\documentclass[12pt,a4paper]{article}
\usepackage{amsmath}
\usepackage[font={footnotesize,it}]{caption}
\usepackage[top=1in, bottom=1in, left=1in, right=1in]{geometry}
\DeclareCaptionStyle{italic}[justification=centering]{labelfont={bf},textfont={it},labelsep=colon}
\usepackage[title]{appendix}
\usepackage{nicematrix}
 
\usepackage{graphicx,psfrag,epsf,textcomp}
\usepackage{lineno}
\usepackage{setspace}
\usepackage{enumerate}
\usepackage[round]{natbib}
\usepackage[hyphens]{url}
\usepackage{xcolor} 
\usepackage{booktabs, bm, paralist,mathpazo,tikz,longtable,microtype}
\usepackage{xcolor,soul,amsfonts,setspace,mathrsfs,mathtools,placeins}
\usepackage{multirow,array}
\usepackage{float}
\usepackage{subcaption}
\usepackage{longtable}

\PassOptionsToPackage{hyphens}{url}
\usepackage[pdftex,colorlinks=true]{hyperref}
\definecolor{darkblue}{rgb}{0,0,.6}
\definecolor{a0}{rgb}{0.0, 0.5, 0.0}
\definecolor{bistre}{rgb}{0.24, 0.17, 0.12}
\definecolor{amethyst}{rgb}{0.6, 0.4, 0.8}
\definecolor{Rcolor}{RGB}{150,160,190}
\definecolor{blush}{rgb}{0.87, 0.36, 0.51}
\definecolor{brightturquoise}{rgb}{0.03, 0.91, 0.87}
\definecolor{burntorange}{rgb}{0.8, 0.33, 0.0}
\hypersetup{citecolor=darkblue,linkcolor=darkblue,urlcolor=darkblue}
\newcommand*\patchAmsMathEnvironmentForLineno[1]{%
  \expandafter\let\csname old#1\expandafter\endcsname\csname #1\endcsname
  \expandafter\let\csname oldend#1\expandafter\endcsname\csname end#1\endcsname
  \renewenvironment{#1}%
     {\linenomath\csname old#1\endcsname}%
     {\csname oldend#1\endcsname\endlinenomath}}% 
\newcommand*\patchBothAmsMathEnvironmentsForLineno[1]{%
  \patchAmsMathEnvironmentForLineno{#1}%
  \patchAmsMathEnvironmentForLineno{#1*}}%

\AtBeginDocument{%
\patchBothAmsMathEnvironmentsForLineno{equation}%
\patchBothAmsMathEnvironmentsForLineno{align}%\
\patchBothAmsMathEnvironmentsForLineno{flalign}%
\patchBothAmsMathEnvironmentsForLineno{alignat}%
\patchBothAmsMathEnvironmentsForLineno{gather}%
\patchBothAmsMathEnvironmentsForLineno{multline}%
}

\newcommand{\blind}{1}

\graphicspath{{images/}}

\newcolumntype{L}[1]{>{\raggedright\let\newline\\\arraybackslash\hspace{0pt}}m{#1}}
\newcolumntype{C}[1]{>{\centering\let\newline\\\arraybackslash\hspace{0pt}}m{#1}}
\newcolumntype{R}[1]{>{\raggedleft\let\newline\\\arraybackslash\hspace{0pt}}m{#1}}

\newsavebox\CBox

\date{}
\usepackage{orcidlink}

\begin{document}

\def\spacingset#1{\renewcommand{\baselinestretch}
{#1}\small\normalsize} \spacingset{1}

\if1\blind
{
\title{\bf Insurance perspectives on climate-related risks}
\author{\normalsize Michelle Dong \orcidlink{0000-0003-1270-4251}\footnote{Corresponding email: zhe.dong@anu.edu.au}\qquad Aaron Bruhn \orcidlink{0000-0003-0053-7978} \qquad Francis K.C. Hui \orcidlink{0000-0003-0765-3533}\\   \normalsize Research School of Finance, Actuarial Studies and Statistics \\      
\normalsize The Australian National University \\
\\
\normalsize Han Lin Shang \orcidlink{0000-0003-1769-6430} \\
\normalsize Department of Actuarial Studies and Business Analytics \\
\normalsize Macquarie University}
\date{}
\maketitle
} \fi

\begin{abstract}
Emerging climate risks are bringing pressing challenges for insurers globally, as they face growing uncertainty in quantifying climate impacts from physical and transition risks. Climate risks are broad and can manifest through short-term changes and events, as well as long-term and sometimes irreversible changes, such as sea level rise and temperature increases. Significant economic changes are inevitable as countries transition and adapt to mitigate climate-related risks. Given the wide range of potential implications for insurers, we conducted a series of interviews with insurance practitioners and academic researchers to explore the diverse landscape of current perspectives on climate-related risks. The interviews considered how insurers can manage emerging climate risks and the role of pricing as a signal of risk. In the context of unknown unknowns for insurers in understanding long-term impacts of climate change, we identify the importance of scenario analysis in informing forward-looking drivers of change and the importance it plays in meeting evolving regulatory requirements.
\vspace{.1in}

\end{abstract}

\spacingset{1.5}
\newpage

\section{Introduction}

Climate-related risks and impacts are broad and multifaceted. A key challenge for insurers is how to understand and make use of evolving and rapidly growing climate-related data and incorporate this within overall risk management. Further complexity arises from whether traditional approaches to risk management and ``business-as-usual" are sufficient to address the swathe of emerging risks relating to climate change. Regulations and disclosure requirements are also evolving rapidly in response to emerging risks. The combination of these challenges and changes presents a unique set of emerging problems for insurers.

This paper aims to shed light on how climate risk is embedded into risk management frameworks and approaches, particularly through the lens of actuaries and other risk management experts within the insurance industry. We note that for life insurers specifically, climate-related risks have largely been concentrated on the asset side to date; however, as life insurance liabilities tend to be long-term in nature, the longer-term physical and transition risk aspects of climate-related risks are highly relevant. This presents a potential gap for life insurers in understanding climate-related risks. Some recent studies have demonstrated potential climate-driven implications for climate-driven life insurers \citep{IAA20, DBS+24}, further motivating the focus of this study.

Through assimilating a range of diverse views, our key research aims are to:
\begin{inparaenum}[1)]
\item distil and identify the most material impacts and risks for life insurers from climate change; 
\item inform how insurance and insurers could adapt in light of these risks; and
\item identify growing areas for future development to help life insurers better understand and manage climate-related risks.
\end{inparaenum}

To address these research objectives, a qualitative approach is adopted through a series of interviews with experts working in industry and academia. An interview-based approach allows us to explore broad perspectives on emerging risks and risk management approaches. This is critical, as research specific to insurance risk management in relation to climate change is nascent, especially with regards to climate data and the quantification of climate-related risks, new risks relating to affordability, protection gaps \& climate migration, and challenges in the development of climate modelling and reporting. This consequently flows into a more specific focus regarding the role of risk experts in insurance, such as actuaries.

We find that the views on the impacts of climate-related risks are broad and varied, which in turn leads to a diversity in the perspectives on risk management. This presents challenges to both academics and practitioners to focus exclusively on one facet without regard for others, and potentially calls for changes in the way risks are currently understood and managed as part of adaptation to new paradigms. The challenges of data reliability and availability also provide the backdrop for growing uncertainty and changing risk. We explore potential actions that insurers can take, including how scenario analysis and modelling can still inform risk management decisions even where data are scarce and uncertainty is high, and how changes in regulation and disclosure requirements could drive further action for insurers. 

The remainder of this paper is structured as follows. In Section~\ref{sec:2} we present a literature review summarising key insights from a range of sources, including academic and industry studies, presentations, public consultations, and government publications. This breadth of sources reflects the significant attention being given to climate change, both in the insurance industry and more broadly. Section~\ref{sec:3} provides an overview of our data and analytical approach, and Section~\ref{sec:4} presents a summary of the interview findings. Section~\ref{sec:5} discusses these findings in more detail, relating them to aspects of current thinking, emerging practice, and industry developments. Section~\ref{sec:6} then concludes. 

\section{Literature Review}\label{sec:2}

Understanding climate-related risks goes beyond the domain of any discipline. It requires a holistic understanding of multiple layers of risk and interactions between them and over time, with a range of implications at the micro and macro levels for individuals, communities, businesses, and governments \citep{AUGOV24}. Within life insurance, the risks and their implications are numerous, interrelated, and significant. Climate-related risks are defined as an insurer’s exposure to physical, transition, and liability risks caused by or related to climate change \citep{IAIS21}. Physical risks generally tend to vary by geographic location and include weather events, as well as longer-term shifts in climate patterns. Transition risks arise when economies transition to a lower carbon economy, bringing about related changes to address mitigation and adaptation to climate change \citep{TCFD17}. Liability risks represent risks around climate litigation, where businesses do not sufficiently respond to climate risks \citep{DBS+24}.

To date, the insurance and actuarial perspective has largely concerned understanding and modelling the implications of damages and the cost of these damages, arising from physical, transition, and liability risks. Physical risk impacts arise from overarching climate-related factors such as increased greenhouse gas emissions, average temperatures and temperature ranges, ocean levels, and rainfall. Responses around adaptation and transitioning to a low-carbon economy then require consideration of transition risk impacts. The combination of physical and transition risks has implications on the economic value of assets and liabilities, and therefore the cost of damages from an insurance perspective \citep{IAA20}. Furthermore, as more data become available for analysis, the paradigm for actuaries and other insurance experts shifts from knowing a little about the impacts to knowing a lot more but still facing great uncertainty. In other words, knowing more does not mean that the inherent uncertainty is any less \citep{M08, M20}. This provides a challenge for insurers, regulators, and consumers in the industry \citep{IAA20}. 

In recent years, governments and regulators across the globe have worked to develop climate risk assessments and climate scenario testing requirements in order to better understand individuals' and companies' exposures to climate risks. One notable example is Australia's first National Climate Risk Assessment \citep[NCRA,][]{ACS25}. This is the first report to deliver a synthesised understanding of the climate risks that Australia is experiencing, and provides insights as to how climate risk affects different sectors and regions of the country. In particular, the key systems Australia's NCRA considers are communities; defence and national security; economy, trade, and finance; health and social support; infrastructure and the built environment; the natural environment, and; primary industries and food. With respect to mortality, the NCRA identifies the current risks around health and social support as moderate, but expects this to evolve into a severe risk by 2050. 

In terms of the financial services sector, the UK Prudential Regulation Authority (PRA) issued a Supervisory Statement (SS5/25) with the intention to enhance the ways in which banks and insurers manage and build resilience against climate-related risks. Notably, one of the key requirements is to advance the use of climate scenario analysis, in order to demonstrate an understanding of how climate scenario outputs inform business decisions \citep{PRA25, PRA25b}.

A key issue for insurers is to understand the concurrent impacts on assets and liabilities \citep{DBS+24}. Testing the balance sheet is an established practice in the industry \citep{LPS110}, which will become more important given the breadth of climate-related risks and their inherent associated uncertainties. From an asset perspective, the focus to date has been on decarbonisation and impact investing \citep{IAA22}. Understanding the liability impacts from climate-related risks has involved studying the causes of loss and drivers of loss \citep{IAA20}, as well as specific factors such as mortality by cause \citep{GGS+17, DSH+24}. A number of scenarios also stress the assets and liabilities, albeit separately, which highlights the need to take a holistic rather than a siloed view on climate risk. 

A comprehensive understanding of both socioeconomic and natural systems is required to determine the implications and impacts of climate-related risks. There are developments globally to more fully consider and disclose climate-related risks and opportunities from a quantitative perspective \citep{CSRD22, ISSB23}. Enhancements and variations of climate scenario narratives including Shared Socioeconomic Pathways (SSPs) have been produced by the Network for Greening the Financial System (NGFS) \citep{NGFS23}, and adapted by central banks and government bodies as part of monitoring and supervision. Examples include the European Insurance and Occupational Pensions Authority (EIOPA) climate scenarios, the Climate Biennial Exploratory Scenario (CBES) testing from the Bank of England, and the Australian scenario testing for banks \citep{BOE22, CFR24}. The role and interpretation of scenario testing and analysis are critical to developing the understanding of climate-related risks, specifically for longer-term risks to which life insurers are exposed.

When it comes to understanding and applying climate scenarios, there is a recognised urgent need for guidance and consistency across government planning, including links to the private sector \citep{ACS25}. The latest draft guidance from the Australian Government on climate scenarios (the National Climate Scenario Guidance, developed by the Australian Government Department of Climate Change, Energy, the Environment and Water) is aimed at supporting climate-related disclosure reporting, climate risk assessment and management, and the design of long-term infrastructure, acknowledging that the understanding and interpretation of climate scenario analysis requires community-wide expertise. Responses to the draft guidance also highlight the significant challenges and diversity of perspectives with respect to managing climate-related risks, ranging from disclosure requirements to operational considerations, but also an overall sentiment that guidance and consistency are important \citep{AI25}. A highly relevant and recent example of uncertainty in the development of appropriate scenarios, along with the need for consistency and robustness, is the recent redaction of a research paper that has directly informed NGFS scenarios \citep{MM25}.

Given the systemic nature of climate risk, a further consideration is the public good or social aspect of insurance. For example, policy decisions to address insurance protection gaps, such as through the Australian Reinsurance Pool (\url{https://arpc.gov.au/}), the UK Flood Re (\url{https://www.floodre.co.uk/}), and the UK Pool Re (\url{https://www.poolre.co.uk/}) are motivated by community and political perspectives. The links between vulnerability and affordability are also being more explicitly considered \citep{PLP+24}. For life insurers, consideration of mortality and morbidity impacts arises from both climate-related physical and transition risks. For example, heightened physical risks may emerge from changes in long-term weather-related trends or through increased frequency and severity of perils. Similarly, the adoption of adaptation measures and changes in the economy to allow for energy transition may also impact mortality and morbidity. 

Finally, with the growing availability of data and the sophistication of  modelling, studies are emerging to improve understanding of the climate-related impacts on risks specific to insurers \citep{DSB22}. However, research into these risks remains nascent, with significant uncertainty regarding how risks will evolve. To the best of our knowledge, this interview-based approach is the first to harvest and distil a wide range of perspectives on climate-related risks for insurers from risk management experts, such as actuaries, to inform how insurance and insurers could adapt, and to identify areas for future development in the face of growing uncertainties and unknown unknowns. 

\section{Data and Methodology}\label{sec:3}

\subsection{Interview approach -- rationale}\label{sec:3.1}

We conducted a series of interviews with leading experts from industry and academia to better understand the key challenges insurers face in identifying and managing climate risks. Interview-based approaches are growing in significance in financial services-related research, partly due to the advantages in collecting perspectives of individuals \citep{Bruhn19, SAD13}.

Interviews provide an opportunity to identify key climate-related issues for practitioners and those garnering attention in academia but are currently underdeveloped in the literature. Capturing this evolving understanding is assisted by interviews taking a more unstructured conversational style. This is because they offer ``maximum flexibility to pursue information in whatever direction appears appropriate, depending on what emerges'' and allows for more opportunities to probe and delve into detail on certain areas \citep{P02}. 

Saturation in interview findings and themes is an important consideration in this approach and is often used as a guide to determine when a sufficient number of interviews have been conducted for data collection \citep{SSK+18, GS17}. In determining the saturation point, we considered other similar qualitative interview approaches and criteria used, including definitions of saturation that depend on the research question \citep{GBJ06}. We also considered the data collected across different stages of our interviews and observed increasingly overlapping themes and perspectives, providing a level of assurance that our interviews were identifying common themes among a number of experts. This practice has been noted to enhance credibility and quality in qualitative research \citep{RK24}. 

\subsection{Data}\label{sec:3.2}

Before approaching prospective interviewees, ethics approval was obtained from the Australian National University in August 2023. %This required careful consideration of a number of issues, including risks involved with the research, incentives for interviewees, the process for gaining interview consent, and confidentiality. 
This addressed research risks and processes around interviewee incentives, consent, and confidentiality. The interviewees were then identified through relevant academic publications or industry presentations, with subsequent referrals to other professionals or academics. This ‘snowball’ feature of obtaining additional referrals is more likely to arise from taking an open-ended, informal interview approach \citep{P02}, and it generated an insightful critique of themes that were emerging from earlier interviews.

%Having first identified some key themes around climate and climate risk management, interviewees were identified and contacted via email based on relevant academic publications or presentations at industry events and conferences. Some interviewees also provided referrals to other professionals or academics, which formed an additional method to identify interviewees. This `snowball’ feature of obtaining referrals is also more forthcoming from taking an open-ended, informal interview approach \citep{P02}, and generated insightful critique on themes that were emerging through earlier interviews. Relevant referrals also helped develop variations and sometimes contrasting views on themes that were emerging through earlier interviews. 

In total, 15 interviews took place between November 2023 and November 2024, with 14 conducted via Microsoft Teams and one face-to-face. The interviews ranged in length from 40 to 75 minutes, and were all digitally transcribed.
 
Eight interviewees were from Industry, four from Academia, two from both, and one from a non-insurance, non-academic background. All industry interviewees were actuaries. Table~\ref{tab:0} shows a summary of the locations and seniority of the interviewees, to provide additional context.

%15 interviews were conducted between November 2023 to November 2024. There were eight interviewees from Industry, four from Academia, two from both, and one from a non-insurance, non-academic background. The ten interviewees from industry were all actuaries. Interviews were conducted via Microsoft Teams, with the exception of one face-to-face interview. Interviews ranged from approximately 40 to 75 minutes. Each interview was digitally transcribed. Table~\ref{tab:0} shows a summary of locations and seniority of the interviewees, to provide context to the discussion topics, responses, and themes identified.

\begin{table}[!htb] 
\caption{Summary of interviewees by location and seniority. Group~2 captures all other interviewees working in climate-related areas.}\label{tab:0} 
\centering
\tabcolsep 0.22in
\begin{tabular}{@{}lll@{}} 
\toprule[1.5pt]
 & \multicolumn{2}{c}{Location} \\ [0.5ex] 
        & Asia-Pacific & Other \\
\cmidrule{1-3}
Group 1: Higher influence roles                     &   & \\
(e.g., professor, CEO, VP, Chief Actuary, Partner)  & 4 & 5 \\
\cmidrule{1-3}
Group 2: All other interviewees                     &   & \\
(e.g,. researchers, actuaries)                      & 3 & 3 \\
\cmidrule{1-3}
\textbf{Total}                                      & \textbf{7} & \textbf{8} \\
\bottomrule[1.5pt]
\end{tabular}
\end{table}

The interviews were conducted on a semi-structured basis to allow the interviewee's broad range of ideas and interests to emerge. Typically, interviews began by discussing climate risk management within insurance, and, as required, prompts were used to focus the discussion on the key research aims.

\subsection{Validation}\label{sec:3.3}

The use of electronic recording of interviews ensured that a record was thoroughly reviewed and that interviewees received timely transcriptions to verify and validate within one week of each interview. This member check provided an important validation tool, with two transcripts returned, one with minor redaction and one with a higher level of redaction, along with further elaboration and clarification. There were no other objections to the completeness or accuracy of the transcripts.
 
%The use of electronic recording of interviews ensured we had a record to review, transcribe, and reflect on more fully after the interview. It helped us more accurately record interview discussions, identify key themes and overall findings, and provide interviewees timely transcriptions for checking and validation. Such member checking is an important validation tool in qualitative research and was predominantly used to ensure the fidelity of their recollection of the conversation, and to add or clarify anything in hindsight. Interviewees were provided with a draft transcript within one week of their interview. Through this process, two transcripts were returned with further elaboration and clarification, one was returned with a minor redaction, and one was returned with a higher level of redactions. There were no other objections to the completeness or accuracy of the transcripts. This relatively low response rate of member checking is not unusual in similar research \citep{DL11}.

The approach to assess what key themes arose from interviews necessarily involved judgement (a key tool of qualitative research \citep{P02}). This judgement was enhanced by having some pre-formed ideas, but not rigid convictions, around some of the likely key issues regarding climate risk management within insurance. For themes that were observed as recurring and therefore significant, we allocated direct quotes from interviews against each theme to highlight specific insights and inform discussion. The inclusion of direct quotes also provides additional validation to the judgement we have applied, as they help maintain the original context of the discussion point \citep{P02}.

%The approach to assess what key themes arose from interviews necessarily involved judgement, which is a key tool of qualitative research \citep{P02}. Themes refer to key factors, relationships, and dependencies between points raised in the data \citep{ABM02}. Identification of these themes then informed our main statements of findings. This judgement was enhanced by having some pre-formed ideas, but not rigid convictions, around some of the likely key issues regarding climate risk management in life insurance.

%For themes that were observed as recurring and therefore significant, we identified and allocated direct quotes from interviews against each theme to help draw specific insights and inform discussion. The inclusion of direct quotes when presenting details of the interview findings also serves as additional validation to the interpretation we have brought to bear on interview material. Direct quotes help with ensuring the interpretation is not misconstrued when summarised into key findings and themes, and maintains the original context of the discussion point. The use of specific quotes also helps to articulate the range of perspectives across the different themes \citep{P02}. 

\section{Interview findings and themes}\label{sec:4}

As introduced in Section~\ref{sec:2}, scenario analysis is quickly developing to be one of the key methods for firms to understand climate-related risks. This sentiment is echoed in interview findings. 

In this section, we first present the full set of findings and related themes from our interviews in Table~\ref{tab:1}. We then focus on one central finding concerning the use of scenario analysis, which, despite challenges, is an approach most interviewees support to better understand the uncertainty posed by climate-related risks. Other major findings include the need for insurance companies to adapt to changes from climate-related risks, the role of insurance pricing, the opportunities for evolving regulations to drive change, and the growing unknown unknowns specific to life insurance risks. These reflect areas of agreement across interviewees, with distinctions between risks and uncertainty discussed at length, as well as the relationship between the two in the context of insurance.

The key findings and related themes within each finding are summarised in Table~\ref{tab:1}.
\begin{longtable}{@{}p{6cm}|p{9.5cm}@{}}
\caption{Summary of findings and related themes from interview discussions}\label{tab:1} \\\toprule
 Finding & Theme \\
 \midrule
Scenario analysis can help actuaries and insurers to understand the range of risks in different ways & Climate-related risks change the ways that models are used and produced \\\cline{2-2}
& Shifts in the distribution impact outcomes and require updates in the analysis more frequently \\
\midrule
Insurance companies need to adapt to changes caused by climate-related risks &  Cost-benefit of being reactive vs leading (first mover disadvantages)      \\
\cline{2-2}
    &  Insurers taking a "wait and see" approach     \\
\midrule
Insurance pricing can be used as a signal to measure climate-related risks for risk management & Insurance pricing as a risk indicator \\\cline{2-2}
 & Insurance pricing driving mitigation and adaptation \\\cline{2-2}
& Insurance pricing and protection gaps \\
\midrule
Regulations and disclosure requirements are developing rapidly, and are a potential lever for driving change & Traditional methods of dealing with events and managing risk are not working anymore \\\cline{2-2}
 & Impact underwriting and impact investing \\
\midrule
Climate-related risks present growing uncertainties and new unknown unknowns & Mortality/morbidity impacts from climate are generally less well understood: climate risk is just one of the risks impacting life insurers \\\cline{2-2}
& Data are required for developing new or enhancing existing tools. Climate data may be an issue; creative approaches and other methods are required \\
\bottomrule
\end{longtable}

\subsection{Scenario analysis}\label{sec:4.1}

An important method to understand the uncertainty presented by climate-related risks is through scenario analysis. Scenarios can help users form a view of the range of risks, wider implications, and systemic interactions. However, in the case of climate, there are limitations in using historical data to inform realistic future scenarios. This is in part due to changes in climate-related trends and the uncertainty around transitioning to a low-carbon economy, meaning that historical analysis is not sufficient to predict future outcomes around insured losses.

As discussed in Section~\ref{sec:2}, recent regulatory guidance encourages insurers to improve the use of scenarios in considering the impact of climate-related risks \citep{PRA25b}. SS5/25 states explicitly that historic data series are not sufficient in themselves to address climate risk, and consequently requires scenario analysis as a key tool for managing climate-related risks. The use of scenarios extends to considerations around materiality of impacts for insurers, as scenarios can help inform insurers of the potential severity and impact of a range of climate outcomes on the insurers' balance sheets.

Six interviewees discussed the increased uncertainty arising from the use of historical data. Whilst acknowledging “\textit{no model is ever correct or right. It never fully predicts the future}” [Interviewee~11], the need to overlay scenarios and data with judgement was highlighted. This is not a new challenge, and neither is it unique to climate-related risks. The interviewees commented on two main themes: firstly, that the uncertainty with climate risk changes the way models are produced and used; and secondly, as climate risk evolves and shifts the distribution and frequency of extreme tail events, a more active control cycle approach is required, including more frequent reviewing and updating key assumptions and modelling approaches. Section~\ref{sec:5} further elaborates on the implications from the outcomes of the interviews on scenario analysis and its challenges.

Table~\ref{tab:5} summarises the key themes relating to scenario analysis.
\begin{longtable}{@{}p{7.5cm}|p{3.7cm}|p{3.7cm}@{}}
\caption{Summary of finding 1 - scenario analysis can help actuaries and insurers to understand the range of risks in different ways}\label{tab:5}
\\\midrule
Grouping of quotes into key statements & Key themes & Finding \\
\midrule
``You have the combination between climate change, the economy, and people’s behaviour… you need (to) set up a very complex model. And as it’s so complex with so many unknown variables, you won’t get it right, that is the problem…” (Interviewee~12) & Climate-related risks change the ways that models are used and produced & Scenario analysis can help actuaries and insurers to understand the range of risks in different ways \\
\cline{1-1}
``It's back to communicating the uncertainties. So, which are the uncertain parameters?… what does that mean for assets that I've got in my book and (for) the key parameters.” (Interviewee~8) & & \\
\cline{1-1}
``It just seems very open-ended because… what are these different scenarios? It could be in terms of a sort of time projection. It could be for the next year, or for the next five years. Or is it on different pathways?”
(Interviewee~6)& & \\
\cline{1-1}
``You don’t replace your modelling with scenario analysis, but you’ve got to become more robust” (Interviewee~2)  & & \\
\cmidrule{1-2}
``If you use the one five years ago and said alright, these are the transition impacts that I'm going to model; those basic things have not panned out anywhere near what was expected” (Interviewee~8) & Shifts in the
distribution impacting outcomes and requiring updates in the analysis more frequently &  \\
\cline{1-1}
``We do not operate in a static system; it is a dynamic system where there are reactions to actions we take. It does not mean that our modelling is useless, and it does not mean in any way that we should not be modelling or doing data science… that stuff is critically important. You’ve got to understand where it has limitations” (Interviewee~2) & & \\
\bottomrule
\end{longtable}

\subsubsection{Climate-related risks change the ways that models are produced and used}\label{sec:4.1.1}

With new unknowns and greater uncertainty from climate-related risks, part of improving the modelling framework will come down to clear communication, and part of it will be driven by ``\textit{the model governance that is set up around it}” [Interviewee 9]. The interviewed practitioners in particular commented that the communication of results to the users also needs to be considered carefully, to ensure there is clarity around ``\textit{the key parameters (and) the wide range of uncertainty around each of them}” [Interviewee 8]. This also highlights the need for a well-engaged Board, which would be responsive to regular and material updates in key modelled outcomes as a way of managing climate risks. 

Multiple interviewees reflected on whether the models are fit for purpose, whether parameter selections remain sensible, and whether limitations around modelled outputs have been clearly communicated to the users of the outputs. Even with these considerations, one interviewee noted that attribution still can present an issue, where ``\textit{sometimes it is really not 100\% clear how much of (the outcome) is really attributable to climate change and what would have happened anyway}” [Interviewee 12].

\subsubsection{Changes in the distribution impacting outcomes}\label{sec:4.4.2}

In addition to taking a more active and responsive approach with modelling and scenario analysis, climate scenarios themselves are becoming increasingly complex. As such, a complex model might be important, but this means that there are “\textit{so many unknown variables that you won’t get it right, that is the problem}” [Interviewee 12]. This, in turn, presents potential issues around model calibration, challenges to model parsimony, and can also be counter-productive to clear communication of results, which, as discussed, is critical. It also highlights the importance of scenario analysis, and through ``\textit{run(ning) different scenarios and see(ing) how things might change}” [Interviewee 12].

The interviewees also noted that the frequency of tail events is expected to increase, along with shifts in the distribution of insured losses from changes in climate-related factors. This has implications for modelling, informing decisions, and presents challenges around “\textit{how you actually shift the whole distribution and how you fatten the tail of the extremes}” [Interviewee 11]. Such changes also make it difficult for users of scenario analysis to interpret and infer what it means for their own risk management. This includes determining appropriate stresses in terms of understanding the ultimate level of uncertainty:  “\textit{models will tell you anything ... based on the inputs and the way that you want to model it. But if there's a bunch of things that you don't know about, a model's not going to tell you that}” [Interviewee 11].

Changes in climate risks shifting the distribution of insurance risks were discussed at length with interviewees, especially practitioners. Interviewee 11 noted a possible upside from climate related changes, with, for example, the potential for a reduction in cold weather mortality as winters become warmer in some regions. In other examples discussed, it was not always clear whether the changes would be positive or negative e.g., changes in transportation routes driven by economies transitioning away from coal production to low-carbon economies.

In the previous section, the critical importance of clear communication was highlighted, and this becomes even more important with the uncertainties from climate risks. This communication is a key part of performing the modelling: ``\textit{the other part about modelling… when you are talking about climate, is the messaging}” [Interviewee 11]. When it comes to scenario modelling, shifts in the distribution of severity and frequency of events impact insured losses, and understanding the business implications of these shifts is an important aspect of managing climate-related risks. As such, interpretation and clear communication are again highlighted as critical, particularly in multidisciplinary settings, where not all stakeholders may understand the outcomes and implications of modelling in detail.

\subsection{Insurance companies and the need to adapt}\label{sec:4.2}

Historically, when new risks have emerged, insurance companies have adapted by changing the way they operate and manage risks, such as through product design and pricing. In this sense, climate-related risks can be considered in a similar way to other risks that have emerged over time. Nearly half of the interviewees expressed the need for active and pragmatic adaptation from insurance companies in response to climate-related risks. Some practitioners emphasised that insurance companies need to adapt more quickly to climate risk, whilst others observed that insurers have already begun to respond in light of growing uncertainty.

This finding is relevant for life and non-life insurers, both of whom will have to consider how climate risk is managed throughout the entire business operation. Indeed, similar to how operational risk can be considered, climate-related risks can impact each management function to varying degrees. For example, the impact on asset liability management (a key risk management tool) may be greater for life insurers who hold long-duration investment assets to match their long-duration liability portfolio, whereas the climate-risk impact on the liability side may be lower than for some non-life insurers due to limited exposure to physical climate effects. To date, life insurers have tended to focus climate-related risk management and adaptation on the asset portfolio.

Table~\ref{tab:2} below summarises the main themes relating to insurance companies needing to adapt.

\begin{longtable}{@{}p{7.5cm}|p{4cm}|p{3.5cm}@{}}
\caption{Summary of finding 2 -- insurance companies need to adapt to changes driven by climate-related risks}\label{tab:2} \\
\toprule
Exemplar quotes leading to relevant theme & Key themes & Finding \\
\midrule
``…(there are a) few instances, (but) less in the climate space, but
more generally in the ESG space around the first mover
disadvantages…” (Interviewee 8) & Cost-benefit of being reactive vs leading (first mover disadvantages) & Insurance companies need to adapt to changes caused by climate-related risks\\
\midrule
``It may emerge in circumstances (where the risk) is going to evolve over time… an insurer does have some capacity to take a “wait and see” approach, and just basically adjust any kind of pricing or ways in which they conduct their business.”
(Interviewee 11)  & Insurers taking a wait and see approach & \\\cline{1-1}
``…it’s a structural (issue) that needs to be resolved. It's not going
to be resolved by one or two companies, (it should) be resolved
collectively.” (Interviewee 7) &  & \\
\bottomrule
\end{longtable}

\subsubsection{Cost-benefit of being reactive vs leading (first mover disadvantages)}

An important point concerns first-mover disadvantages for insurers who act and adapt to growing and emerging risks. Interviewee 8 highlighted that first mover disadvantages may be material for insurers who react to growing risks through pricing, or making investment choices away from carbon-intensive assets. The context of the competitive landscape is therefore an important factor in an insurer’s planning and options for adaptation. There is a perception that this needs to be balanced with meaningful acknowledgement and management of climate-related risks.

\subsubsection{Insurers taking a "wait-and-see" approach}

Conversely, some practitioners see insurers as being slow to react, and have taken more of a ``wait-and-see'' approach, especially in light of how uncertain and systemic climate-related risks are. This tends to be where the cost of acting sooner outweighs the benefit of pre-emptively acting. However, it was also highlighted that such an approach would lead to greater risks compared to acting sooner: “\textit{sometimes it takes a health scare before someone starts to take care of their well-being... (people) have a natural tendency to ignore things that they cannot see or feel. And that's probably (going to) happen with climate as well}” [Interviewee~11]. Practitioners and academics alike acknowledged the risks that come with adopting a ``wait-and-see” approach: “\textit{one of the big problems we…have (is) to make it understandable for people and acceptable for people (to realise) that we have to act now}” [Interviewee~12]. The general consensus from interviewees is that insurers will need to change the way they react to climate risk, and whilst some insurers have begun to do so, there is more that can be done.

The interviews also highlighted that insurers are not the only agents with the ability to drive adaptation to climate-related risks. This leads to further discussions on the catalyst effect from changes in regulations (Section~\ref{sec:4.5}). How governments, regulators, and individuals in society also play a role in shaping the trajectory more broadly are also explored further as part of Section~\ref{sec:5}.

\subsection{Insurance pricing as a risk signal}\label{sec:4.2}

In setting premiums, insurers have the potential to use pricing as a risk signal to governments, individuals, and businesses. Generally, insurers have more information on the collective risks: they are in a unique position to quantify systemic risks (such as climate) in a more considered way than the typical individual or business. For example, an individual’s exposure to climate risk will be specific to their own circumstances, whereas an insurer’s portfolio could have a different overall risk profile. Accurately pricing for these risks can be a way of signalling to those who have less information. 

Insurance pricing was discussed at length with seven interviewees, and highlighted that price is a mechanism to drive and motivate individuals, businesses, and governments to mitigate climate risks. A point more relevant to general rather than life insurance concerned the motivation for large-scale adaptation measures.

Table~\ref{tab:3} below summarises the main themes relating to insurance pricing.

\begin{longtable}{@{}p{8.2cm}|p{3.5cm}|p{3.5cm}@{}}
\caption{Summary of finding 3 - insurance pricing as a signal for climate-risk mitigation or adaptation.}\label{tab:3} \\
\toprule
Grouping of quotes into key statements & Key themes & Finding \\
\midrule
``Insurers… have a very critical role, particularly in price
signaling.” (Interviewee 11) & Insurance pricing as a risk indicator & Insurance pricing can be used as a signal to measure climate-related risks for risk management \\
\cline{1-1}
``One of the puzzles (for) the insurance sector is how to know how to price mitigation efforts. So at the moment… There is a very loose link between any resilience or mitigation done to a home, and a change in pricing to insurance…” (Interviewee 5) & &  \\
\cmidrule{1-2}
``An example is where you’ve established areas that you can build and live in…(and) to influence decisions and say, your premium is X here, but it is 4X over there… an insurer has a real (role) to play in influencing behaviour beyond what
governments in the public sector do.” (Interviewee 11) & Insurance pricing driving mitigation and adaptation & \\
\cmidrule{1-1}
``People who are poor are affected much more than people who are rich. So normally rich people, they can really quite easily manage climate change for themselves…”
(Interviewee 12) & & \\
\cmidrule{1-2}
``…The problem really is that these prices might not be
affordable for the public any longer.” (Interviewee 12) & Insurance pricing and
protection gaps & \\
\cline{1-1}
``…You don't want to make it too expensive, because then
insurance becomes unaffordable, and people don't buy it.”
(Interviewee 6) & & \\
\cline{1-1}
``The affordability problem was not caused by climate change,
and not caused by bad weather. It was caused by the actuaries
applying technology and data” (Interviewee 2) & & \\
\bottomrule
\end{longtable}

\subsubsection{Insurance pricing as a risk indicator}\label{sec:4.3.1}

Insurers play a unique role in price signalling by being able to set prices that reflect the risk and uncertainty of the insured event. The interviewees reflected on the importance and difficulty of pricing accuracy, as pricing needs to take into account factors that concern ``\textit{what is going to happen in 20 years time}” [Interviewee 4]. The expectations for insurers to price signal were explored in depth across several interviews, for example: “\textit{insurers, in particular, have a very critical role, particularly in price signalling}” [Interviewee 11]. As a risk indicator, insurance prices are an early warning signal to others in the market and can help identify new and emerging issues. The interviewees recognised that climate is more than just an insurance issue, that it “\textit{is a societal issue}” [Interviewee~11], and that “\textit{society as a whole… (is) probably another angle to look at}” [Interviewee 7].

\subsubsection{Insurance pricing driving mitigation and adaptation}\label{sec:4.3.2}

Insurance pricing can drive behavioural reactions from those faced with paying insurance premiums, including mitigating actions. As an example, insurance premiums becoming prohibitively expensive in certain areas could drive people living in those areas to relocate to places less exposed to climate risks. This is how insurance pricing can be used to “\textit{persuade (policyholders) to move to a safer area}” and where “\textit{an insurer has a real clear advantage and a real role to play in influencing behaviour}” [Interviewee 11]. However, quantifying the benefit of such mitigation efforts does not come without challenges, with one interviewee commenting that “\textit{at the moment, generally speaking, there is a very loose link between any resilience or mitigation done to a home, and a change in pricing to insurance}” [Interviewee 5]. This can lead to some risk mitigation efforts not being accurately reflected in pricing in the short term.

The ability to price the signal also comes with considerations on insurance affordability and protection gaps, which were discussed at length with four interviewees. Ultimately, for climate-related risks, insurance pricing can be useful to reflect risk and uncertainty, which can motivate adaptation and mitigation measures from individuals and society. In extreme cases where insurance becomes unaffordable, insurance pricing could also be considered an indication of the need for government intervention and protection gap entities.

\subsection{Regulation and disclosures}\label{sec:4.4}

One view expressed across interviewees was that the link between climate impacts, financial risk, and regulatory capital is tenuous. However, it was also noted that it may change as regulations and disclosure requirements evolve. The rationale for decisions driven by disclosures also tends to differ depending on whether they are regulatory-driven responses or driven by public pressure, with five interviewees commenting on differences between the current climate-related regulation being perceived as a compliance exercise versus driving meaningful change.

Six interviewees commented on regulations and disclosures, with all noting that although requirements differ slightly by jurisdiction, they are nevertheless developing rapidly, and represent a potential lever for driving longer-term change. 

Table~\ref{tab:6} below summarises key themes related to this finding.
\begin{longtable}{@{}p{7.5cm}|p{3.5cm}|p{4cm}@{}}
\caption{Summary of finding 4 - regulations and disclosures are developing rapidly, and are a potential lever for driving change}\label{tab:6}\\
\toprule
Grouping of quotes into key statements & Key themes & Finding \\
\midrule
``Somehow it's trying to fit all the regulations. So you need to have
profitability, you need to have competitive prices, and you need to be
compliant. So you try to be everything.” (Interviewee 4) & Traditional
methods of dealing with events and managing risk not working anymore & Regulations and disclosure requirements are developing rapidly, and are a potential lever for driving change\\
\cline{1-1}
``…[reporting is] not really transparent because it's very complicated and you still have lots of different ways of interpreting data. You have a lot of freedom and you need this lot of freedom because companies are different, and industries are different, so it is hard to compare… but in
the end, you want to compare” (Interviewee~12) & & \\
\cmidrule{1-2}
``There are no capital implications. The ISSB and TCFD and others they are just reporting standards… it's a good start is measure your emissions…But it's not penalising or incentivising you. ” (Interviewee~8)  & Impact underwriting and impact investing & \\
\cline{1-1}
``Once you can prove that there is a connection between climate risk and financial stability… I think it will be easier for regulators to bring in this sort of rules that either incentivise impact underwriting or penalise
underwriting of, you know, coal and oil and gas.” (Interviewee~6) & & \\
\bottomrule
\end{longtable}

\subsubsection{Traditional methods of dealing with events and managing risk are not working anymore}\label{sec:4.4.1}

Recent developments in climate regulation, including the International Sustainability Standards Board (ISSB) and the Corporate Sustainability Reporting Directive (CSRD), are primarily aimed at transparency and helping insurers understand risk drivers and exposures. Interviewee~11 commented that understanding where the risks are arising will help policymakers and governments make decisions around resilience and intervention: ``\textit{It’s an important thing to get more transparency and understanding… where that risk sits in the system… You can either strengthen your resilience, or you can just recognise who's going to cop it and whether they have a choice in it, whether they're aware of it, and whether they decide to actually do something about it or not.}'' However, with these changes in regulation, interviewees who were also practitioners in life insurance reflected whether the changes are merely a compliance exercise for insurers, or whether these changes can drive real changes in managing climate risk and improving climate resilience. An interviewee stated that these new regulations and disclosures have no impact. Rather, “\textit{the biggest effect, at least on the short term, is really that companies get more transparency for themselves. What we currently don't see... is real transparency on an aggregate level because all these analyses, ratings, or all the reporting, that's so heterogeneous}” [Interviewee 12]. Whilst increased transparency for insurers is a beneficial first step for insurers, there remains a way to go to improve transparency across different insurers and other firms. In a similar vein, another interviewee noted that “\textit{there are no capital implications (from the new regulations). The ISSB and TCFD (are) reporting standards … it's a good start (to) measure your emissions (and) hopefully that brings some transparency around what kind of company you are, but it's not penalising or incentivising (firms)}” [Interviewee 8].

Although things are evolving through regulation, much remains to be done in terms of incentivising real action. This was highlighted by interviewee 8, who noted that regulations are ``\textit{probably better than nothing, but there is an aspect of “well this is a compliance exercise and I'm just going to do it because I'm being told to do it” rather than any particular change of behaviour}”. Another interviewee noted that “\textit{the standards people are ahead of the actuarial people}” [Interviewee 2], whereas another noted that the “\textit{regulator is not doing anything about (climate risk)}” [Interviewee 4]. So whilst there is a shift towards transparency in helping insurers to better manage climate-related risks, the general observation from interviewees is that there are still changes required to fully enable transparency and eventually drive meaningful changes in behaviour. 

\subsubsection{Impact underwriting and impact investing}\label{sec:4.4.2}

As regulations change, impact underwriting and impact investing are expected to become more important for life insurers. Impact underwriting refers to the ability for insurers to contribute to adaptation through underwriting practices in terms of data, risk assessment, and expertise \citep{EIOPA23}. Impact investing is defined as the deployment of funds into investments that generate a measurable social or environmental impact alongside a financial return on investment \citep{GIIN17}. Both impact underwriting and impact investing are important for life insurers to consider in the context of their liability portfolio and typically long-duration assets.

Although interviewees observed that current climate-related regulations are largely around disclosure and transparency, ultimately, this has the potential to drive changes that manifest through impact underwriting and impact investing. Both practitioners and academics expressed a range of views on responsible investing for life insurers, and that changes and developments in climate regulation could drive changes in behaviour and drive insurers to act. For example, responsible investment “\textit{seems to be a thing that's occupying a lot of some of the commercial companies at the moment because they have to do it}” [Interviewee 8]. Various interviewees observed that different insurers’ responses depended on company-specific factors, relating to the earlier discussion around the need to adapt (Section~\ref{sec:4.1}). To limit first-mover risks, there is an impetus for regulation to drive consistency in actions. For example; ``\textit{this is why you need the regulation, because all the laggards don't want to do anything}” [Interviewee 8]. Notably, in the Australian context, there is a strong link between the life insurance industry and the superannuation industry, and the size of the latter is such that the scale of their investment and societal expectations around the energy transition is significant\footnote{Group insurance schemes provide insurance cover to related groups of people, for example, members of a superannuation fund that predominantly services select industries. The large size of group insurance and the superannuation industry in Australia tends to generate more attention from the public as changes have greater impact on the collective funds and insurance cover.}. As a result, “\textit{society is expecting people, the firms, or the powers that be to move quicker, the politicians to move faster}” [Interviewee 8]. Therefore, insurance companies working with these large superannuation firms will also need to adapt, driven by societal expectations and pressure.

By setting and disclosing net zero targets, insurers are also faced with reputation and regulatory risk if they fail to meet and comply with their own net zero targets. Interviewee 11 acknowledged this and commented that “\textit{there's a disincentive (to disclose)… a lot of the time (insurers) want to water that down and say (that) a portion of our investments is green or in socially acceptable investments}”. Therefore, additional disclosure requirements could become a barrier for impact investing, as firms are currently not incentivised or rewarded for their green investment choices. 

Overall, there was a view that although the understanding of climate risks remains nascent, regulations and disclosure requirements for life insurers will likely evolve as uncertainties are better understood, and regulations shift to more explicitly “\textit{drive behaviours}” [Interviewee 8]. 

\subsection{Growing uncertainties and unknown unknowns}\label{sec:4.5}

In the shifting landscape of growing uncertainty arising from climate-related risks, interviewees also discussed some specific considerations for life insurers. Two themes emerged, and were discussed at length with five interviewees: mortality and morbidity impacts from climate-related risks are less well understood; and that better and deeper data are ultimately required for actuaries and others working in life insurance to develop and enhance existing tools to help understand climate-related risks better.

Table~\ref{tab:7} summarises relevant quotes relating to each of the three themes discussed.
\begin{longtable}{@{}p{6.5cm}|p{4.5cm}|p{4cm}@{}}
\caption{Summary of finding 5 - climate-related risks present growing uncertainties and new unknown unknowns}\label{tab:7} \\
\toprule
Grouping of quotes into key statements & Key themes & Finding \\
\midrule
``What we observe… more or less worldwide, is that the effects of climate change are really minor compared to many of the other effects we currently observe” (Interviewee 12). & Mortality/morbidity impacts from climate are generally less well understood: climate risk is just one of the risks impacting life insurers & Climate-related risks present growing uncertainties and new unknown unknowns\\
\cline{1-1}
``There is a liability side impact for life insurance as well. But it's very difficult to quantify. And so it seems to be… a little bit brushed under the carpet, those sort of liability potential liability side risks because they're a bit second order” (Interviewee 8).& &  \\
\cmidrule{1-2}
``Sometimes too much data, and sometimes not enough data. There are parts of the industry where it is a lot more developed… but then other areas (that have) less and less granular (and) less available (data).” (Interviewee 9). & Data are required for developing new or enhancing existing tools. Climate data might be an issue, therefore other methods balancing reliance on data with judgement are required & \\
\cline{1-1}
``It's a very wide cone of uncertainty… you've got quite a number of variables that could swing quite substantially during that transition period… it feels like (all) the predictions from five years ago have been blown out of the water and it's much worse. So you get the latest IPCC report and you start to base your modelling on that.” (Interviewee 8) & & \\
\cline{1-1}
``We can look at the past (and) torture it until it confesses… run it through 15 GLM models and all this data science and we come up with an answer. And that’s the way we have always done things. But in the future you are going to
have to learn a different skill, which is scenario analysis.” (Interviewee 2) & & \\
\cline{1-1}
``You don’t throw all your models in the trash, because your models are a framework to help you think about the problem. So clearly understanding what the models are telling you is fine, but you can’t come up with stresses from your models based on the past when your future does not look like the
past” (Interviewee 2) &  & \\
\bottomrule
\end{longtable}

\subsubsection{Mortality/morbidity impacts from climate are generally less well understood}\label{sec:4.5.1}

To date, life insurers have largely focused on transition risk. This is driven by the long-duration assets that life insurers typically hold to back liabilities, and the relatively greater exposure to transition risk, compared to the slower trends that impact life insurance liability portfolios. Practitioner interviewees, therefore, noted a potential gap for life insurers in understanding mortality and morbidity impacts from climate-related risks. For example, “\textit{there is a liability side impact for life insurers as well. But it is very difficult to quantify… because they are a bit second order}” [Interviewee~8].

The interviewees acknowledged that other emerging issues, such as the broader economy, developments in healthcare, longevity, and public policy, are all critical factors an insurer needs to consider as well. Trends in climate-related physical and transition risks tend to be slower-moving, with longer-term impacts, so many life insurers have placed more emphasis on more immediately pressing challenges. Interviewee 12 noted that “\textit{what we observe really, more or less worldwide, is that the effects of climate change are really minor compared to many of the other effects (life insurers) currently observe. For instance, in North America, issues like obesity, the opioid crisis, (have) a bigger effect on mortality/morbidity than climate change}” [Interviewee~12].

Even if life insurers prioritise understanding the mortality and morbidity impact from climate-related risks, there are additional challenges with attributing the impact on mortality or morbidity to climate risk. This makes it more difficult for life insurers to quantify specific impacts from climate-related risks and to adopt mitigating actions. Overall however, interviewees were in agreement that climate-related risks add to the existing risks facing life insurers, with one interviewee hypothesising that \textit{“climate change will perhaps also in the future not be the main impact}” [Interviewee 12], and another stating that “\textit{climate change is no different, it is just like that, it’s like air bags in cars, or legalising cannabis… it’s going to affect your losses}” [Interviewee 2].

\subsubsection{There are limitations, but data are required for developing new or enhancing existing tools}\label{sec:4.5.2}

Limitations around data add to the uncertainty for life insurers to understand climate-related risks. Interviewees expressed concerns with overly relying on data-driven methods, particularly as the “\textit{cone of uncertainty}'' [Interviewee 8] becomes wider and wider. Five interviewees commented that life insurers should balance reliance on historical data with an element of judgement. These data limitations: (``\textit{a spectrum of sometimes too much data, and sometimes not enough data}'' [Interviewee 9]) can subsequently hinder the ability to draw strong conclusions, which in turn means more emphasis on the application of judgement from experts such as actuaries. This view was shared by all interviewees who discussed data limitations. Interviewee~12 noted that ``\textit{all these modelling exercises, all this scenario building, is really complex and you need a lot of judgement... You have to consider a lot of different expert opinions and they might even contradict each other, because we have not yet observed it in our data, and so it is really hard to build models to project into a future that has not been observed before.}'' Furthermore, interviewee 8 commented that “\textit{data is always difficult. You'll never get perfect data}'', and interviewee 2 stated that “\textit{actuaries need to stop being data scientists}” and move away from an over-reliance on data alone.

Interviewees acknowledged that even uncertain historical data is still useful for developing new tools and/or enhancing existing tools. This is central to any good practice, and particularly to developing the understanding of emerging climate-related risks. Interviewee 2 articulated that actuarial models “\textit{are a framework to help you think about the problem}” and clarified “\textit{you do not throw all your models in the trash}”, simply because the future trends may not be fully captured in historical data. However, there is a clear limitation, “\textit{understanding what the models are telling you is fine, but you can’t come up with stresses from your models based on the past when your future does not look like the past}”.

Therefore, actuaries and life insurers need to consider more creative approaches and methods to consider the implications of climate-related risks: “\textit{what we can do is still data-driven, but based on models produced by scientists and their observations. Again, the future reality nobody can really know, especially when it comes to people. Nobody knows how people will behave. So especially for life and health ... that is going to be quite difficult to assess}” [Interviewee 12]. This reiterates the overall uncertainty in dealing with climate-related risks.

\section{Discussion}\label{sec:5}

From the five key findings, the need for insurers to proactively understand and manage emerging climate risks is clear. To help better understand and manage climate-related risks, there was a clear appetite to focus on how scenario analysis can inform decisions, and how regulations can enhance transparency and be a catalyst for change. 

Scenario analysis is a key area with potential to enhance our overall understanding and implications of climate-related risks. This stems from wider research used to understand climate risks, including the IPCC shared socioeconomic pathways and representative concentration pathways. However, since the ``funnel of uncertainty" is significant in the context of climate risk, the way actuarial models are produced and used may be different for the use of scenarios in other actuarial models and applications. For instance, shifts in the distribution where the extreme events happen with more frequency (and severity) likely mean that models require more frequent updating and reassessment, as part of ongoing risk management. %To date, life insurers have focused more on understanding the implications of climate transition risk on the asset side of the balance sheet, as they often hold long-duration assets to match the long-duration nature of their liabilities, therefore exposing them heavily to climate transition risk. For liabilities, the natural hedge between protection and longevity portfolios is another avenue to understand and quantify through scenario analysis.

Recent regulatory guidance recognises the limitations of solely relying on historic data to address and manage climate risk, with some guidance requiring the use of scenario analysis. Materiality plays a major role in considering climate scenario analysis, to help inform insurers of the potential severity and impact of a range of climate outcomes on an insurer’s balance sheet.
%Recent regulatory guidance recognises the limitations to solely relying on historic data to address and manage climate risk, with some guidance requiring the use of scenario analysis in considering the impact of climate-related risks. Materiality plays a major role in considering climate scenario analysis, to help inform insurers of the potential severity and impact of a range of climate outcomes on the insurers’ balance sheets.
Notwithstanding this, the use of scenario analysis is an area that requires careful consideration and ongoing monitoring as our understanding of climate risks evolves and grows. In most cases, the choice of scenarios to adopt requires tailoring to the risk exposures of the insurer and materiality. The choice of scenario also requires constant updating and monitoring, as evidenced by the recent redaction of certain NGFS scenarios upon further review \citep{MM25}.

In relation to considerations around materiality, the PRA also acknowledges the requirement for proportionality when it comes to choices around climate scenarios. This is especially important for insurers where exposure to physical and transition risks may vary, leading to more sophisticated scenario analysis for the more likely and material risks \citep{PRA25b}. For life insurers specifically, we observe that there has been more focus on assets and investments relative to the liability exposures. This is partly due to the limited physical risk implications for life insurance portfolios, in combination with the potential for greater transition risk impacts arising from the longer duration assets that life insurers typically hold to match their liabilities. Furthermore, for liabilities, the natural hedge between protection and longevity portfolios is another avenue to understand and quantify through scenario analysis. One further and important factor is the time horizon: the intended application of scenarios can mean long-term scenarios are considered, for example, the NGFS scenarios based on IPCC pathways, or have a more tailored and bespoke view of short-term risks through short-term scenario modelling. 

Other findings present broader views from interviews and implications for insurers. One finding is that life insurers will have to develop options to adapt to growing climate risk and uncertainty. The balance between the costs and benefits of taking a wait-and-see approach compared to being a first mover requires careful consideration from insurers, which can be informed by case studies and relevant examples from first movers in the past. 

A parallel discussion with interviewees was cyber risk, which represents a similarly emerging risk in the recent past. Initially, a number of insurers introduced exclusions to remove the uncertainty around cyber risk on their portfolios, as part of adopting more of a ``wait-and-see" approach. Over time, as the industry's understanding of this risk increased, insurers started removing the exclusions. A critique of such use of exclusions under a "wait-and-see" approach related to the obvious protection gaps that resulted, although others considered the uncertainty was unquantifiable enough to justify alternative courses of action.

Yet another finding highlights the role of insurance pricing as a signal of risk. One response might be to price for the uncertainty arising from anticipated increases in claims, though this would have obvious competitive disadvantages. A more nuanced approach will have to balance market forces, the role of insurance as a pooling mechanism versus rating specific risks in detail, and broader pricing inputs such as impacts on policy lapses, economic assumptions, and changes in any approaches to underwriting. Further consideration of specific life insurance product features, distribution channels and platforms, and the target demographic and exposure to climate risk across this demographic, would also be essential. Put simply, the combined effect of all such influences on pricing will require deeper reflection on the emerging risks from climate change, and a consideration of overall business goals in light of the emerging risks.

The life insurance industry is heavily regulated in most parts of the world, and given the long-term nature of risks insured and duties to policyholders, the role of regulation and societal expectations becomes even more important in this regard, to both reflect and shape desired behaviours. Currently, there is a regulatory focus in many countries on disclosure and ensuring transparency for investors and the public. Interviewees acknowledged that the biggest effect of regulations to date is in bringing about greater transparency for insurers, but that comparisons across insurers and other firms remain a challenge. Some regulations can be more akin to a compliance exercise, rather than driving meaningful change through impact underwriting and impact investing. This is an important point because, similar to research into climate-related risks, regulations are also at a nascent stage. Whilst improved transparency is a first step and beneficial for insurers, further developments could also have the potential to become a catalyst for change. 

There are data limitations and gaps for life insurers in the collective understanding of mortality and morbidity impacts from climate-related risks. In light of such uncertainty and potentially unreliable data, actuarial judgement becomes more important, as data analytics based on historical data will have its limitations when the past is less relevant to inform the future. Nevertheless, whilst climate-related risks are systemic and impact almost all aspects of risk, they are also just one of many risks that life insurers need to consider when making decisions around pricing, product design, and how to adapt to changes. 

\section{Conclusion}\label{sec:6}

A range of views and opinions was obtained from a series of interviews with practitioners and academic researchers in respect of climate risks for insurers, and life insurers in particular. These views highlight key aspects of climate-related risks that can help inform both future research, and industry considerations. Importantly, the consensus of views is that climate-related risk is not a niche topic, but is an emerging systemic risk with highly uncertain outcomes, that must be understood and addressed through a range of perspectives and skills, all the while dealing with data limitations and consequent need for significant judgement. The future in managing these risks for life insurers requires meaningful collaboration between actuaries, practitioners, and researchers. Furthermore, it also requires the leveraging and ongoing development of risk management tools, including long-term projections and scenario analysis.

The use of interviews has strengths in bringing out the views of others, especially when the researcher and process are open to what the important issues are, rather than having a fixed, pre-conceived view. This is important in the context of climate-related risks, where different views do exist. However, interviews can also give rise to insights that are less clear, where statements are ambiguous and/or contradictory. Nevertheless, this does not prevent the key findings from being drawn out, as has been done in this paper. There are also potential biases introduced as a result of interpretation by the interviewer and the perspectives of the interviewees themselves. Extending the interviews to a wider group of participants may provide other perspectives that may be equally insightful in understanding climate-related risks for life insurers, which is a future goal of this research.

Over time, as climate risk and efforts to address it evolve, ongoing research can also consider the effectiveness or otherwise of mitigation efforts. This will be a critical insight to help insurers and others better meet the challenges of an uncertain future.

\section*{Acknowledgment}

The authors are grateful for the comments by two reviewers and the participants at the Climate Change and Insurance workshop at the Vienna University of Technology in 2024, and thank the participants at a seminar at Macquarie University for their comments and questions. We also thank the interviewees for their interest, time, and contributions to this research.

\newpage
\bibliographystyle{agsm}
\bibliography{cci2024}

\end{document}